\documentclass[aps,pra,twocolumn,reprint,noeprint,superscriptaddress,amsmath,amssymb,floatfix,%
]{revtex4-2}

\usepackage[T1]{fontenc}
\usepackage[utf8]{inputenc}
\usepackage{amsmath,amssymb,amsfonts,bm,mathtools}
\usepackage{graphicx}
\usepackage{hyperref}
\usepackage{orcidlink}
\usepackage{xcolor}
\usepackage{booktabs}
\usepackage{physics}
\usepackage{braket}
\usepackage{float}
\hypersetup{
    colorlinks=true,
    linkcolor=blue,
    citecolor=blue,
    urlcolor=blue}

\makeatletter
\newcommand{\cofirstmark}{%
  \appdef\@affil@present@script{\textsuperscript{,\,\hyperlink{frontmatter.1}{*}}}%
}
\makeatother

\begin{document}

\title{Quantum-enhanced multiphase estimation with spin-1 Bose--Einstein condensates}

\author{Jiaheng Jin}
\thanks{These two authors contributed equally to this work.}
\affiliation{Beijing Academy of Quantum Information Sciences, Beijing 100193, China}
\affiliation{State Key Laboratory of Low Dimensional Quantum Physics, Department of Physics, Tsinghua University, Beijing 100084, China}
\author{Xinwei Li~\orcidlink{0009-0005-9677-0011}\texorpdfstring{\cofirstmark}{}}
\email[Contact author: ]{xinwei.li1991@outlook.com}
\affiliation{Beijing Academy of Quantum Information Sciences, Beijing 100193, China}
\author{Dong E. Liu~\orcidlink{0000-0002-1295-2362}}
\affiliation{Beijing Academy of Quantum Information Sciences, Beijing 100193, China}
\affiliation{State Key Laboratory of Low Dimensional Quantum Physics, Department of Physics, Tsinghua University, Beijing 100084, China}
\affiliation{Frontier Science Center for Quantum Information, Beijing, China}
\affiliation{Hefei National Laboratory, Hefei, Anhui 230088, China}
\author{Li You~\orcidlink{0000-0002-3919-4768}}
\email[Contact author: ]{lyou@mail.tsinghua.edu.cn}
\affiliation{Beijing Academy of Quantum Information Sciences, Beijing 100193, China}
\affiliation{State Key Laboratory of Low Dimensional Quantum Physics, Department of Physics, Tsinghua University, Beijing 100084, China}
\affiliation{Frontier Science Center for Quantum Information, Beijing, China}
\affiliation{Hefei National Laboratory, Hefei, Anhui 230088, China}
\date{\today}

\begin{abstract}
We investigate quantum-enhanced two-phase estimation in a spin-$1$ Bose--Einstein condensate, where spin-exchange collisions generate a spin-nematic squeezed state in the $m_F=\pm1$ modes.
The resulting pair-correlated state has a rank-one quantum Fisher information matrix (QFIM) for the two mode phases relative to the $m_F=0$ mode and therefore cannot resolve them independently.
Nevertheless, we find that a parameter-independent passive three-mode transformation lifts this rank deficiency and enables joint estimation of both phases.
In the undepleted-pump regime, we derive expressions for the QFIM, optimize the equal-weight multiparameter quantum Cram\'er--Rao bound, and obtain a full-rank probe with squeezing-enhanced sensitivity below the matched fixed-total-number standard quantum limit (SQL).
These analytical predictions are confirmed by exact-diagonalization calculations.
We further construct a parameter-independent readout that maps the phase combinations onto displacement channels accessible through atomic homodyne or parity measurements.
Our results provide a route from spin-mixing-generated quantum correlations to two-phase estimation in spinor condensates.
\end{abstract}

\maketitle

\section{Introduction}
\label{sec:intro}
Quantum metrology exploits nonclassical resources, such as squeezed and entangled states, to achieve measurement precision beyond classical limits~\cite{Giovannetti2011,Pezze2018,Degen2017}.
The quantum Fisher information provides a general framework for quantifying the sensitivity of a quantum state to an encoded parameter and determines the corresponding quantum Cram\'er--Rao bound~\cite{Braunstein1994,Helstrom1976,Holevo1982,Paris2009}.
While quantum-enhanced sensing is well established for single-parameter estimation, many applications require several parameters to be estimated simultaneously, including multiphase interferometry, vector-field sensing, and distributed quantum sensing~\cite{Humphreys2013,Baumgratz2016,Kasevich2022,Treutlein2026}.
Multiparameter quantum estimation is naturally described by the quantum Fisher information matrix (QFIM), which defines the corresponding multiparameter quantum Cram\'er--Rao bound (QCRB)~\cite{Szczykulska2016,Liu2019,Gessner2018,Ragy2016,Albarelli2020}.
For pure-state unitary models with commuting generators, the QCRB can be jointly saturated in local estimation and therefore gives the attainable precision limit~\cite{Ragy2016,Pezze2017}.
Optimizing a multiparameter sensor thus requires consideration of the full information matrix rather than optimizing each parameter separately.

Bose--Einstein condensates provide a versatile setting for coherent many-body dynamics and quantum-enhanced interferometry, ranging from Josephson dynamics to squeezed-state sensing~\cite{Smerzi1997,Gross2010,Riedel2010,Klempt2011,Linnemann2016}.
Spinor condensates offer additional internal degrees of freedom and provide a particularly useful platform for generating nonclassical states~\cite{Ho1998,Ohmi1998,Law1998,Kawaguchi2012,StamperKurn2013,Oberthaler2008}.
For instance, in a spin-$1$ condensate, spin-exchange collisions coherently convert pairs of atoms in the macroscopically occupied $m_F=0$ component into correlated atoms in the $m_F=\pm1$ components~\cite{Chang2005,Kronjager2005,Klempt2010}.
In the undepleted-pump regime, this dynamics is described by two-mode squeezing and produces strong pair correlations, including nearly perfect number-difference correlations between the $m_F=+1$ and $m_F=-1$ modes~\cite{mustecapliouglu2002spin,Klempt2010,Hamley2012}.
At the same time, the three Zeeman components provide a natural setting for multiphase sensing: relative to the $m_F=0$ component, the phases accumulated by the $m_F=\pm1$ components contain common and differential contributions associated with quadratic- and linear-Zeeman-like perturbations, respectively.
Such phase shifts can arise from magnetic fields, microwave or optical dressing, and vector or tensor light shifts~\cite{Vengalattore2007,Gerbier2006,Santos2007}, while related multilevel phase shifts are also relevant to quantum sensing and tests of fundamental physics~\cite{Zhuang2018Spinor, Li2019LSV, Ma2026}.

The pair-correlated state generated by spin mixing is not by itself a complete two-parameter probe.
For the encoding $\exp[-i(\theta_{+}\hat N_{+1}+\theta_{-}\hat N_{-1})]$, it responds only to the common phase combination $\theta_{+}+\theta_{-}$, while remaining insensitive to the differential phase combination $\theta_{+}-\theta_{-}$, so that the QFIM has rank one.
Such spin-mixing systems have already been widely explored for quantum-enhanced sensing.
One important route in spin-$1$ condensates is matter-wave $SU(1,1)$ interferometry, which realizes the $SU(1,1)$ framework of nonlinear interferometry with spin-mixing dynamics~\cite{Yurke1986,Jing2011,Chekhova2016,Manceau2017,Szigeti2017PumpedUp,Opanchuk2012,Oberthaler2011}.
Beyond $SU(1,1)$ interferometry, spinor condensates also support quantum-enhanced sensing based on entanglement and cyclic dynamics~\cite{Luo2017,Zou2018,Mao2023,liu2022nonlinear}, including the simultaneous estimation of orthogonal spin rotations~\cite{Cao2025JointEstimation}.
The present task is different: we seek to simultaneously estimate two relative phases generated by the commuting operators $\hat N_{+1}$ and $\hat N_{-1}$.
The central challenge is therefore to convert this pair-correlated resource into a full-rank two-phase probe using experimentally accessible linear operations while enabling practical readout of both phases.

In this work, we develop a quantum-enhanced two-phase estimation protocol based on the spin-nematic squeezed state generated by spin-exchange collisions in a spin-$1$ Bose--Einstein condensate.
We optimize the probe preparation over passive three-mode transformations and identify a compact two-step operation that converts the initially rank-one phase response into two independent phase-sensitive directions.
The transformation reorganizes the pair correlations and couples the macroscopically occupied $m_F=0$ component to both sensing channels, thus allowing the squeezing resource to enhance the simultaneous estimation of the two phases beyond the standard quantum limit (SQL).
In the undepleted-pump regime, we obtain analytical expressions for the QFIM, the phase-matching condition, and the optimal mixing angle, yielding a full-rank two-parameter probe with a simple physical implementation.
Numerical calculations quantitatively reproduce the analytical optimum and confirm the predicted sub-SQL regime.
We further construct a parameter-independent readout that maps the two encoded phase responses onto measurable displacement channels accessible by atomic homodyne or parity detection.
These results provide an experimentally feasible route from spin-mixing-generated squeezing to simultaneous quantum-enhanced estimation of two relative phases in a spinor condensate.

The remainder of the paper is organized as follows.
In Sec.~\ref{sec:model}, we introduce the spin-mixing dynamics and the two-phase estimation protocol.
In Sec.~\ref{sec:qfim}, we derive and optimize the QFIM and compare the analytical predictions with numerical calculations.
In Sec.~\ref{sec: readout}, we construct the phase-to-displacement mapping and analyze the atomic homodyne and parity readout strategies.
Finally, Sec.~\ref{sec:conclusion} summarizes the main results.

\section{Physical model and estimation protocol}
\label{sec:model}
\begin{figure}[htbp]
    \centering 
    \includegraphics[width=1\columnwidth]{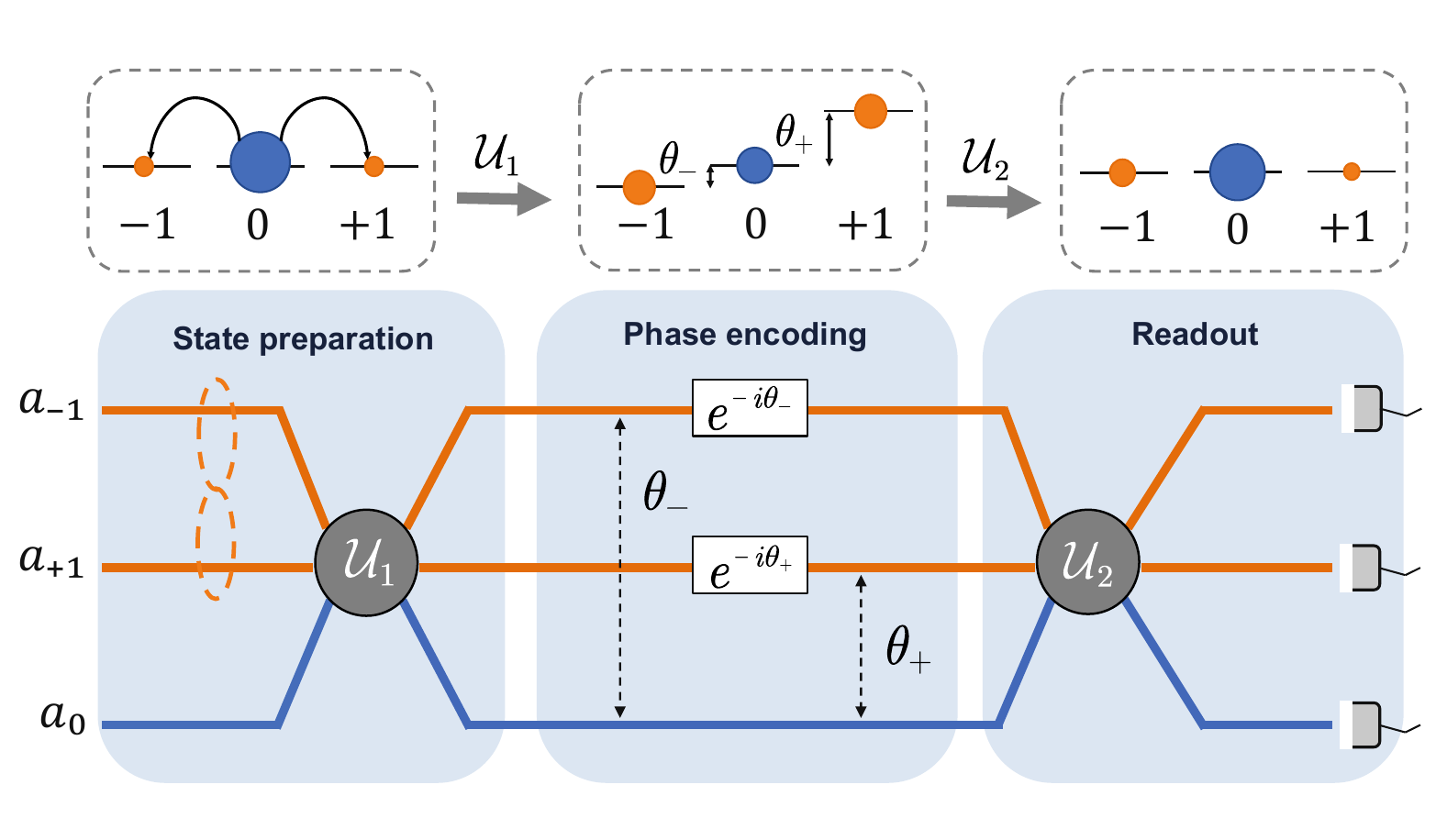}
    \caption{Schematic of quantum-enhanced multiphase estimation with a spin-$1$ BEC. Spin-mixing dynamics prepares a spin-nematic squeezed state by generating correlated atom pairs in the $m_F=\pm1$ modes from the initially populated $m_F=0$ mode. The first three-mode mixing operation $\mathcal U_1$ transforms this state into a probe sensitive to two independent phase combinations. The phases $\theta_{-}$ and $\theta_{+}$ are encoded on the $m_F=-1$ and $m_F=+1$ modes, respectively, relative to the $m_F=0$ reference. The second mixing operation $\mathcal U_2$ recombines the three modes and maps the encoded phase information onto measurable output observables.} \label{fig:protocol} 
\end{figure}

\subsection{Spin-mixing Dynamics in Spin-1 BEC}
We consider a spin-$1$ BEC in the single-mode approximation (SMA), in which all three Zeeman components share the same normalized spatial wave function,
$\hat{\Psi}_{m}(\mathbf r)=\phi(\mathbf r)\hat a_m$ with $m=0,\pm1$.
We work in a sector of fixed total atom number $N_t$, so that $\sum_m\hat N_m=N_t$, where $\hat a_m$ and $\hat a_m^\dagger$ are the annihilation and creation operators of mode $m$, and $\hat N_m=\hat a_m^\dagger\hat a_m$.
After omitting the spin-independent contribution, which is constant for fixed $N_t$, the spin-dependent Hamiltonian can be written as~\cite{Law1998,Kawaguchi2012,StamperKurn2013}
\begin{equation}
\begin{aligned}
\hat H=&\frac{c_2}{2N_t}\bigg[2\left(
\hat a_{+1}^{\dagger}\hat a_{-1}^{\dagger}\hat a_0\hat a_0
+\hat a_{+1}\hat a_{-1}\hat a_0^{\dagger}\hat a_0^{\dagger}
\right)
\\
&\quad+\left(\hat N_{+1}-\hat N_{-1}\right)^2
+\left(2\hat N_0-1\right)\left(\hat N_{+1}+\hat N_{-1}\right)\bigg]
\\
&-p\left(\hat N_{+1}-\hat N_{-1}\right)+q(\hat N_{+1}+\hat N_{-1}).
\label{eq:SMA_H_full}
\end{aligned}
\end{equation}
Here, the spin-dependent interaction strength is $c_2=N_t\lambda_a\int d^3r\,|\phi(\mathbf r)|^4$, with $\lambda_a=4\pi\hbar^2(a_2-a_0)/(3M)$, where $M$ is the atomic mass and $a_0$ and $a_2$ are the $s$-wave scattering lengths in the total-spin $0$ and $2$ channels.
The parameters $p$ and $q$ denote the linear and quadratic Zeeman shifts, respectively.
The pair-conversion term in Eq.~\eqref{eq:SMA_H_full} coherently converts two atoms in the $m_F=0$ mode into one atom in each of the $m_F=\pm1$ modes, and vice versa.

We assume that the condensate is initially prepared in the polar state
\begin{equation}
|\psi(0)\rangle
=
|0,N_t,0\rangle ,
\label{eq:polar_initial_state}
\end{equation}
where the three entries denote the occupations of the $m_F=+1,0,-1$ modes, respectively, and all atoms occupy the $m_F=0$ mode.
Because spin-exchange collisions conserve the longitudinal magnetization, the evolution remains in the zero-magnetization subspace and the linear Zeeman term does not affect the preparation dynamics.
In the short-time, weak-depletion regime, the $m_F=0$ mode can be treated as an undepleted pump.
Taking its phase as the reference, we use $\hat a_0,\hat a_0^\dagger\rightarrow\sqrt{N_t}$.
For the ferromagnetic case considered here, $c_2<0$, tuning the quadratic Zeeman shift to $q\simeq |c_2|$ cancels the leading side-mode number term.
Up to an additive constant and subleading depletion corrections, the resulting Hamiltonian reduces to the two-mode squeezing form
\begin{equation}
\hat H_{\rm sm}
=
-|c_2|
\left(
\hat a_{+1}^{\dagger}\hat a_{-1}^{\dagger}
+
\hat a_{+1}\hat a_{-1}
\right),
\label{eq:H_sq}
\end{equation}
which describes the coherent creation and annihilation of correlated pairs in the initially empty $m_F=\pm1$ modes, driven by the macroscopically occupied $m_F=0$ pump~\cite{Klempt2010,Hamley2012}.

The dynamics generated by Eq.~\eqref{eq:H_sq} is conveniently expressed through the Heisenberg evolution of the $m_F=\pm1$ mode operators.
Introducing the dimensionless squeezing parameter $r=|c_2|t$ and setting $\hbar=1$, we obtain
\begin{equation}
\hat a_{\pm1}(t)
=
\hat a_{\pm1}(0)\,\cosh r
+
i\hat a_{\mp1}^{\dagger}(0)\,\sinh r.
\label{eq:a_pm_Bog}
\end{equation}
Acting on the initially empty side modes, this Bogoliubov transformation produces, within the undepleted-pump approximation, the two-mode squeezed-vacuum state
\begin{equation}
|\psi_{\rm sq}^{(\pm)}\rangle
\simeq
\frac{1}{\cosh r}
\sum_{k=0}^{\infty}
\left(i\tanh r\right)^k
|k\rangle_{+1}\otimes |k\rangle_{-1}.
\label{eq:TMSV_pair_state}
\end{equation}
Here, $|\psi_{\rm sq}^{(\pm)}\rangle$ denotes the approximate state of the $m_F=\pm1$ sector obtained by treating the $m_F=0$ condensate as undepleted.
The underlying three-mode spin-mixing dynamics, however, remains number conserving.

Within the undepleted-pump description, the pair-correlated state in Eq.~\eqref{eq:TMSV_pair_state} represents the spin-nematic squeezing generated by spin-mixing dynamics in the spin-$1$ condensate~\cite{mustecapliouglu2002spin,Hamley2012}.
For this state, the two $m_F=\pm1$ modes have identical mean populations and number fluctuations,
\begin{subequations}
\label{eq:population_statistics}
\begin{align}
\langle \hat N_{+1}\rangle
=
\langle \hat N_{-1}\rangle
&=
\sinh^2 r ,
\label{eq:population_mean}
\\
(\Delta \hat N_{+1})^2
=
(\Delta \hat N_{-1})^2
&=
\sinh^2 r \cosh^2 r .
\label{eq:population_variance}
\end{align}
\end{subequations}
Because atoms are created in correlated pairs, the state also satisfies
\begin{equation}
\Delta^2(\hat N_{+1}-\hat N_{-1})=0 ,
\label{eq:number_difference_correlation}
\end{equation}
which expresses perfect number-difference correlations between the $m_F=+1$ and $m_F=-1$ modes.
As shown below, the absence of fluctuations in the differential number operator makes the direct two-phase response rank one and motivates the pre-encoding mode transformation.

\subsection{Multiphase Interferometric Protocol}
After the spin-mixing dynamics has prepared the spin-nematic squeezed resource, we consider the simultaneous estimation of two phases accumulated by the $m_F=\pm1$ modes relative to the $m_F=0$ reference.
Taking the $m_F=0$ phase as the reference and omitting an irrelevant global phase, the encoding is described by
\begin{equation}
U_{\boldsymbol{\theta}}
=
\exp\left[
-i\left(
\theta_{+}\hat N_{+1}
+
\theta_{-}\hat N_{-1}
\right)
\right],
\end{equation}
where $\boldsymbol{\theta}=(\theta_+,\theta_-)^{\mathsf T}$, and $\theta_+$ and $\theta_-$ are the relative phases accumulated by the $m_F=+1$ and $m_F=-1$ components, respectively.

During the interrogation time $T$, we assume that the evolution is generated by the mode-dependent shifts alone.
During this stage, $p$ and $q$ denote the effective differential and common shifts, respectively.
Using the sign convention of Eq.~\eqref{eq:SMA_H_full}, the accumulated phases are
\begin{equation}
\theta_{+}
=
(q-p)T,
\qquad
\theta_{-}
=
(q+p)T .
\end{equation}
The differential component $p$ shifts the two $m_F=\pm1$ modes in opposite directions, whereas the common component $q$ shifts them together relative to the $m_F=0$ reference.
Consequently, $\theta_{+}+\theta_{-}=2qT$ and $\theta_{-}-\theta_{+}=2pT$ encode the common and differential phase components, respectively.
For magnetic field sensing, these correspond to quadratic- and linear-Zeeman-related contributions, although the same parametrization also applies to mode-dependent phases generated by other physical perturbations.

Although the two phases are encoded on the $m_F=\pm1$ modes, Eq.~\eqref{eq:number_difference_correlation} shows that the spin-mixing-generated state has no fluctuations in the differential number operator.
Consequently, direct phase encoding produces a rank-one response and does not constitute a complete two-parameter probe.
We therefore introduce a parameter-independent passive three-mode transformation $\mathcal U_1$ before phase encoding and subsequently optimize it to obtain two independent phase-sensitive directions.
The complete interferometric sequence is shown schematically in Fig.~\ref{fig:protocol} and is written as
\begin{equation}
|\psi_{\boldsymbol{\theta}}^{\rm out}\rangle
=
\mathcal U_2
U_{\boldsymbol{\theta}}
\mathcal U_1
|\psi_{\rm sq}\rangle ,
\end{equation}
where $|\psi_{\rm sq}\rangle$ denotes the full three-mode state produced by the spin-mixing preparation stage, whose undepleted-pump side-mode representation is given in Eq.~\eqref{eq:TMSV_pair_state}.
The transformation $\mathcal U_1$ prepares the phase-sensitive probe before encoding, whereas the parameter-independent transformation $\mathcal U_2$ is applied after phase encoding.
Because $\mathcal U_2$ is unitary and independent of $\boldsymbol{\theta}$, it does not change the QFIM of the encoded state; instead, it maps the phase response onto experimentally accessible readout channels.
These channels are connected to spin-resolved detection through the atomic homodyne and parity measurement strategies developed below.

\section{Optimization of the Quantum Fisher Information Matrix by Linear Transformation}
\label{sec:qfim}
Having specified the spin-nematic squeezing generation and the two-phase encoding protocol, we optimize the passive three-mode transformation $\mathcal U_1$ before phase encoding. Its single-particle mode matrix belongs to $U(3)$; for fixed total atom number $N_t$, an overall mode phase contributes only a global phase to the many-body state. We therefore use an $SU(3)$ representative and formulate the optimization in terms of the QFIM of the encoded probe state.

\subsection{Definition of Quantum Fisher Information Matrix and the Sensitivity Bound}\label{subsec: optimizing_qfim}
For a fixed pre-encoding passive three-mode transformation $\mathcal U_1$, the phase-encoded probe state is
\begin{equation}
|\psi_{\boldsymbol{\theta}}\rangle =
e^{-i(\theta_{+}\hat N_{+1}+\theta_{-}\hat N_{-1})}
\mathcal U_1|\psi_{\rm sq}\rangle .
\end{equation}
For this pure-state family, the corresponding quantum Fisher information matrix (QFIM) is
\begin{equation}
(F_Q)_{ij}=4{\rm Re}
\left[
\langle \partial_i\psi_{\boldsymbol{\theta}}|\partial_j\psi_{\boldsymbol{\theta}}\rangle
-
\langle \partial_i\psi_{\boldsymbol{\theta}}|\psi_{\boldsymbol{\theta}}\rangle
\langle \psi_{\boldsymbol{\theta}}|\partial_j\psi_{\boldsymbol{\theta}}\rangle
\right],
\label{eq:qfim}
\end{equation}
where $\partial_i\equiv\partial/\partial\theta_i$ and $i,j\in\{+,-\}$.

For direct phase encoding without pre-encoding mode mixing, $\mathcal U_1=\mathbb I$. Within the undepleted-pump approximation, Eqs.~\eqref{eq:TMSV_pair_state} and~\eqref{eq:qfim} give the rank-one QFIM
\begin{equation}
F_Q
=
4\sinh^2 r\cosh^2 r
\begin{pmatrix}
1 & 1\\
1 & 1
\end{pmatrix}.
\end{equation}
The nonzero and null eigendirections of this matrix correspond to the common and differential phase directions, respectively. This rank deficiency follows from the number-difference constraint $(\hat N_{+1}-\hat N_{-1})|\psi_{\rm sq}\rangle=0$, which makes the differential phase inaccessible. Consequently, the two phases cannot be jointly identified from the bare probe. Moreover, the bare generators $\hat N_{+1}$ and $\hat N_{-1}$ act only on the side modes, so the macroscopically occupied $m_F=0$ pump does not directly contribute to the phase-sensitive fluctuations. The pre-encoding transformation $\mathcal U_1$ is therefore designed both to lift this rank deficiency and to coherently mix the pump mode into the effective phase generators.

Equivalently, the pre-encoding transformation can be transferred to the phase generators acting on the initial squeezed state. The resulting effective generators are
\begin{equation}
\hat G_{\pm}(\mathcal U_1)
=
\mathcal U_1^\dagger \hat N_{\pm1}\mathcal U_1 .
\label{eq:eff_generator}
\end{equation}
These operators are the side-mode number generators pulled back through $\mathcal U_1$ to the input state. For unitary encoding of a pure state, the QFIM is four times the covariance matrix of these effective generators:
\begin{equation}
F_Q=
\begin{pmatrix}
A & C\\
C & B
\end{pmatrix},
\end{equation}
where $A = 4 \text{Var}[\hat{G}_{+}(\mathcal{U}_1)]$, $B = 4 \text{Var}[\hat{G}_{-}(\mathcal{U}_1)]$, and $C = 4 \text{Cov}[\hat{G}_{+}(\mathcal{U}_1), \hat{G}_{-}(\mathcal{U}_1)]$, with all moments evaluated in $|\psi_{\rm sq}\rangle$. Here, $A$ and $B$ are the diagonal quantum-information contributions associated with the two phase coordinates, whereas $C$ quantifies their covariance-induced coupling. Thus, varying $\mathcal U_1$ changes the full generator covariance matrix and hence the joint precision bound.

For $\nu$ independent repetitions, the multiparameter quantum Cram\'er--Rao bound (QCRB) constrains the covariance matrix of any locally unbiased estimator $\widehat{\boldsymbol{\theta}}$ as
\begin{equation}
{\rm Cov}(\widehat{\boldsymbol{\theta}})
\succeq
\frac{1}{\nu}F_Q^{-1}.
\end{equation}
Because the effective generators commute, the pure-state compatibility condition is satisfied, and the matrix bound provides a jointly attainable benchmark in the local-estimation setting. In what follows, we quote the bound per interrogation, corresponding to $\nu=1$. For a full-rank QFIM and equal weights on the two phases, we use the QCRB on the sum of the estimation variances, $\operatorname{Var}(\widehat{\theta}_{+})+\operatorname{Var}(\widehat{\theta}_{-})$, as the scalar cost,
\begin{equation}
\Delta\Theta^2_{\rm QCRB}
=
{\rm Tr}(F_Q^{-1})
=
\frac{A+B}{AB-C^2}.
\label{eq:trace_inverse_qfim}
\end{equation}
Here, invertibility requires $AB-C^2>0$. A singular QFIM leaves at least one phase combination unidentifiable and therefore corresponds to an infinite two-parameter cost. The optimization of $\mathcal U_1$ thus amounts to minimizing $(A+B)/(AB-C^2)$ over the allowed passive transformations by balancing the generator variances and their cross-covariance. In the following subsection, we evaluate these quantities for a general passive three-mode transformation.

\subsection{Optimization of the QCRB}
\begin{figure}[t]
\centering
\includegraphics[width=0.95\columnwidth]{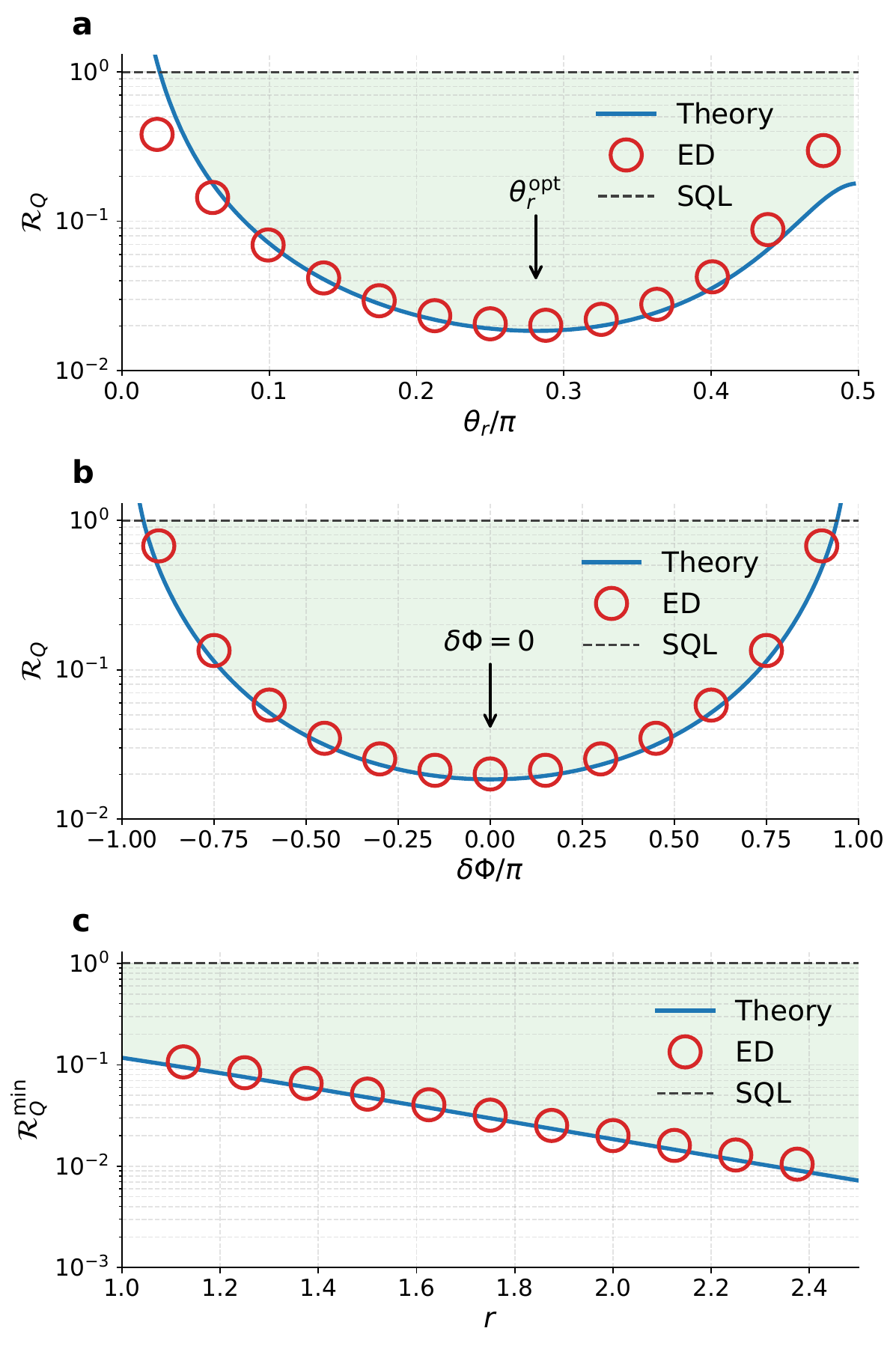}
\caption{
Optimization of the pre-encoding transformation $\mathcal U_1$
for $N_t=1000$.
Blue solid curves (Theory) show the leading-order undepleted-pump
predictions, while red open circles (ED) show fixed-$N_t$
exact-diagonalization results for the spin-1 system.
The metrological gain is expressed by
$\mathcal R_Q=\Delta\Theta_{\rm QCRB}^2/\Delta\Theta_{\rm SQL}^2$,
so that $\mathcal R_Q<1$ corresponds to performance below the SQL.
(a) $\mathcal R_Q$ as a function of the mixing angle $\theta_r$
at $r=2$ under phase matching.
The solid curve is given by Eq.~\eqref{eq:qcrb_Uprime},
and the arrow marks the analytical optimum $\theta_r^{\rm opt}$.
(b) Dependence on the phase mismatch
$\delta\Phi=2\phi_r+\phi_1-\pi/2$
at $r=2$ and $\theta_r=\theta_r^{\rm opt}$.
The arrow marks phase matching at $\delta\Phi=0$.
(c) Minimum SQL-normalized QCRB,
$\mathcal R_Q^{\min}=(\Delta\Theta_{\rm QCRB}^2)_{\min}
/\Delta\Theta_{\rm SQL}^2$,
as a function of the squeezing parameter $r$.
For each $r$, the minimum is obtained by optimizing $\theta_r$
at phase matching; the analytical minimum is given by
Eq.~\eqref{eq:minimum_qcrb_large_r}.
Horizontal dashed lines mark the SQL, and the shaded areas
between the theoretical curves and the SQL highlight
sub-SQL performance.
}
\label{fig:qcrb_optimization}
\end{figure}

The rank-one response identified above shows that spin-nematic squeezing alone does not provide a full-rank resource for estimating two independent phases. We therefore jointly optimize the parameter-independent pre-encoding transformation $\mathcal U_1$ over the eight-parameter $SU(3)$ representatives defined above. A general representative is parameterized as
\begin{align}
\mathcal U_1={}&D(\delta_1,\delta_2)R_{0,-1}(\theta_3,\phi_3)
R_{0,+1}(\theta_2,\phi_2)R_{+1,-1}(\theta_1,\phi_1),
\label{eq:su3_factorization}
\end{align}
where $D(\delta_1,\delta_2)$ applies the phases $-(\delta_1+\delta_2)$, $\delta_1$, and $\delta_2$ to the $m_F=0,+1,-1$ modes, respectively. Each $R_{i,j}(\theta_k,\phi_k)$ is a two-mode Rabi mixing between modes $i$ and $j$, with $\theta_k$ determining the mixing strength and $\phi_k$ the relative coupling phase. Thus, $\boldsymbol{\lambda}=(\delta_1,\delta_2,\theta_1,\phi_1,\theta_2,\phi_2,\theta_3,\phi_3)$ spans all eight real degrees of freedom of the passive $SU(3)$ transformation. For each $\boldsymbol{\lambda}$, we construct the effective generators in Eq.~\eqref{eq:eff_generator} and evaluate the QFIM from the covariance relation in Sec.~\ref{subsec: optimizing_qfim}. We then jointly minimize the equal-weight cost $\Delta\Theta_{\mathrm{QCRB}}^2=\operatorname{Tr}(F_Q^{-1})$ over all eight parameters. The analytical expressions retain the leading terms in the undepleted-pump regime $2\sinh^2r\ll N_t$; with the pump phase chosen as reference, we define $\mu_0>0$ by
\begin{equation}
\mu_0^2\equiv\langle\hat N_0\rangle\simeq N_t-2\sinh^2r.
\label{eq:pump_population}
\end{equation}
The explicit mode-matrix elements, transformed-mode expressions, and moment expansion are given in Appendices~\ref{app:su3_elements} and~\ref{app:noise_stats}. 

The full eight-parameter minimization can be reduced analytically to a one-dimensional problem. Because $D(\delta_1,\delta_2)$ commutes with the number generators $\hat N_{\pm1}$, the phases $\delta_1$ and $\delta_2$ do not affect the QFIM and are therefore not fixed by the optimization. An optimal choice of the remaining parameters satisfies
\begin{subequations}
\label{eq:optimal_SU3_constraints}
\begin{align}
\theta_1
&=\frac{\pi}{4},
\label{eq:optimal_SU3_constraint_a}\\
\sin\theta_2
&=\sin\theta_3\cos\theta_2,
\label{eq:optimal_SU3_constraint_b}\\
-\phi_1+2\phi_3
&=-\frac{\pi}{2}\pmod{2\pi},
\label{eq:optimal_SU3_constraint_c}\\
\phi_1+2\phi_2
&=\frac{\pi}{2}\pmod{2\pi}.
\label{eq:optimal_SU3_constraint_d}
\end{align}
\end{subequations}
Equation~\eqref{eq:optimal_SU3_constraint_a} fixes a balanced rotation between the two side modes, while Eq.~\eqref{eq:optimal_SU3_constraint_b} equalizes the pump-coupling amplitudes of the two sensing channels. Equations~\eqref{eq:optimal_SU3_constraint_c} and~\eqref{eq:optimal_SU3_constraint_d} select the phase combinations for which the pair correlations generated by spin mixing contribute constructively to both sensing channels. The angular constraint in Eq.~\eqref{eq:optimal_SU3_constraint_b} leaves only one independent pump--side-mode mixing degree of freedom. We parameterize this common mixing amplitude by an angle $\theta_r$. The three mixing angles are related by
\begin{equation}
\begin{aligned}
\sin\theta_2
&=\frac{\sin\theta_r}{\sqrt{2}},\\
\sin\theta_3
&=\tan\theta_2
=\frac{\sin\theta_r}
{\sqrt{1+\cos^2\theta_r}}.
\end{aligned}
\label{eq:two_step_mixing}
\end{equation}
The corresponding symmetric pump--side-mode coupling can be implemented by a collective spin-$1$ rotation $R_r(\theta_r,\phi_r)$, where $\theta_r$ determines the mixing strength and $\phi_r$ determines the coupling phase. Defining $c=\cos(\theta_r/2)$ and $s=\sin(\theta_r/2)$, its single-particle matrix in the ordered basis $(m_F=0,+1,-1)$ is
\begin{equation}
R_r(\theta_r,\phi_r)=
\begin{pmatrix}
\cos\theta_r
&-\sqrt{2}cs\,e^{i\phi_r}
&\sqrt{2}cs\,e^{-i\phi_r}\\
\sqrt{2}cs\,e^{-i\phi_r}
&c^2
&s^2e^{-2i\phi_r}\\
-\sqrt{2}cs\,e^{i\phi_r}
&s^2e^{2i\phi_r}
&c^2
\end{pmatrix}.
\label{eq:Rr_definition}
\end{equation}

The same constrained QFIM can therefore be generated by the two-step representative
\begin{equation}
\mathcal U_1
=
R_r(\theta_r,\phi_r)
R_{+1,-1}\!\left(\frac{\pi}{4},\phi_1\right).
\label{eq:structured_U}
\end{equation}
The balanced side-mode rotation reorganizes the pair correlations into two single-mode squeezing directions, after which $R_r$ couples the pump mode symmetrically to both directions. Within this representation, the effect of the two phase constraints in Eqs.~\eqref{eq:optimal_SU3_constraint_c} and~\eqref{eq:optimal_SU3_constraint_d} is reproduced by the single phase-matching condition
\begin{equation}
2\phi_r+\phi_1
=
\frac{\pi}{2}\pmod{2\pi}.
\label{eq:phase_matching}
\end{equation}
Once this condition is imposed, $\theta_r$ is the only sensitivity-relevant parameter that remains undetermined. Minimizing the resulting QCRB below gives $\theta_r^{\mathrm{opt}}$.

To quantify deviations from the phase-matching condition in Eq.~\eqref{eq:phase_matching}, we define the residual phase mismatch $\delta\Phi\equiv
2\phi_r+\phi_1-\pi/2$ and its associated correlation factor as 
\begin{equation}
\chi(\delta\Phi)\equiv
\cosh 2r+\cos(\delta\Phi)\sinh 2r.
\label{eq:chi_definition}
\end{equation}
For the two-step transformation in Eq.~\eqref{eq:structured_U}, retaining the leading $\mu_0^2$ terms in the undepleted-pump regime gives
\begin{subequations}
\label{eq:ABC_simplified}
\begin{align}
A = B &\simeq 16\mu_0^2c^4s^4
      + 8\mu_0^2c^2s^2(c^4+s^4)\chi(\delta\Phi),\\
C     &\simeq 16\mu_0^2c^4s^4
      - 16\mu_0^2c^4s^4\chi(\delta\Phi).
\end{align}
\end{subequations}
The equality $A=B$ reflects the exchange symmetry of the two sensing channels. The QFIM eigenvalues in the differential and common phase directions are
$A-C\simeq8\mu_0^2c^2s^2\chi(\delta\Phi)$ and
$A+C\simeq8\mu_0^2c^2s^2[4c^2s^2+(c^2-s^2)^2\chi(\delta\Phi)]$, respectively.
Thus, nonzero three-mode mixing ($cs\neq0$) produces a full-rank QFIM. At phase matching, $\chi(0)=e^{2r}$, which maximizes the squeezing-enhanced contribution to the QFIM.
Under the phase matching condition, substituting Eq.~\eqref{eq:ABC_simplified} into Eq.~\eqref{eq:trace_inverse_qfim} gives
\begin{equation}
\Delta\Theta_{\mathrm{QCRB}}^2(\theta_r,r)
=
\frac{2c^2s^2+(c^4+s^4)e^{2r}}
{4\mu_0^2c^2s^2e^{2r}
\left[4c^2s^2+(c^2-s^2)^2e^{2r}\right]}.
\label{eq:qcrb_Uprime}
\end{equation}
On the large-squeezing branch, minimization of Eq.~\eqref{eq:qcrb_Uprime} over $\theta_r$ gives
\begin{equation}
\theta_r^{\mathrm{opt}}
=\arcsin\!\left[\sqrt{\frac{(2-\sqrt{2})e^{2r}}
{e^{2r}-1}}\right].
\label{eq:optimal_angle_large_r}
\end{equation}
The resulting minimum is
\begin{equation}
\left(\Delta\Theta_{\mathrm{QCRB}}^2\right)_{\min}
=
\frac{3+2\sqrt2}{2\mu_0^2}
\left(e^{-2r}-e^{-4r}\right).
\label{eq:minimum_qcrb_large_r}
\end{equation}
For $r\gg1$ within the undepleted-pump regime, this result exhibits the leading scaling $\Delta\Theta_{\mathrm{QCRB}}^2\propto e^{-2r}/\mu_0^2$, showing that squeezing lowers the joint variance bound while the pump population sets its overall $1/\mu_0^2$ scale.

To quantify the metrological gain, we compare the QCRB with the SQL for unentangled probes with the same two encoded phases and the same equal-weight cost~\cite{Li2020MultiParameter,Goldberg2020Multiphase},
\begin{equation}
\Delta\Theta_{\mathrm{SQL}}^2
=
\frac{3/2+\sqrt{2}}{N_t}.
\label{eq:sql}
\end{equation}
We define their ratio as
\begin{equation}
\mathcal R_Q
\equiv
\frac{\Delta\Theta_{\mathrm{QCRB}}^2}
{\Delta\Theta_{\mathrm{SQL}}^2}.
\label{eq:qcrb_sql_ratio}
\end{equation}
Thus, $\mathcal R_Q<1$ means that the equal-weight QCRB lies below the SQL.

Figure~\ref{fig:qcrb_optimization} compares the analytical predictions with exact-diagonalization results for the spin-1 system. The numerical calculations reproduce both the optimal mixing angle and the phase-matching condition. They also follow the predicted dependence of the optimized QCRB on the squeezing strength, supporting the validity of the analytical description over the parameter range considered. 
Importantly, these results show that the metrological advantage is not tied to a finely tuned operating point: the optimum in the mixing angle is relatively broad, while the phase dependence remains centered at the predicted matching condition. 
We therefore use the two-step transformation in Eq.~\eqref{eq:structured_U}, evaluated at phase matching and at $\theta_r=\theta_r^{\rm opt}$, as the optimized probe preparation in the remainder of the paper. Having established the QFIM-optimal preparation, we now turn to the measurement problem and construct a parameter-independent readout $\mathcal U_2$ that maps the two encoded phase responses onto experimentally accessible observables.

\section{Readout strategies for simultaneous phase estimation}
\label{sec: readout}
\subsection{Mapping Phases to Quadrature Displacements}
Having fixed the probe preparation in Sec.~\ref{sec:qfim}, we now specify the unitary operation applied after phase encoding and before detection. Here we choose
\begin{equation}
\mathcal U_2=R_r^\dagger(\theta_r,\phi_r).
\label{eq:readout_rotation}
\end{equation}
This inverse rotation recombines the pump and side modes after phase encoding. The final three-mode state is therefore
\begin{equation}
\begin{aligned}
|\psi_{\boldsymbol\theta}^{\mathrm{out}}\rangle
&=\mathcal U_2 U_{\boldsymbol\theta}\mathcal U_1
|\psi_{\mathrm{sq}}\rangle\\
&=R_r^\dagger U_{\boldsymbol\theta}R_r
R_{+1,-1}\!\left(\frac{\pi}{4},\phi_1\right)
|\psi_{\mathrm{sq}}\rangle,
\end{aligned}
\label{eq:readout_output_state}
\end{equation}
where the arguments $(\theta_r,\phi_r)$ of $R_r$ have been suppressed in the second line.

\begin{figure}[t]
\centering
\includegraphics[width=0.95\columnwidth]{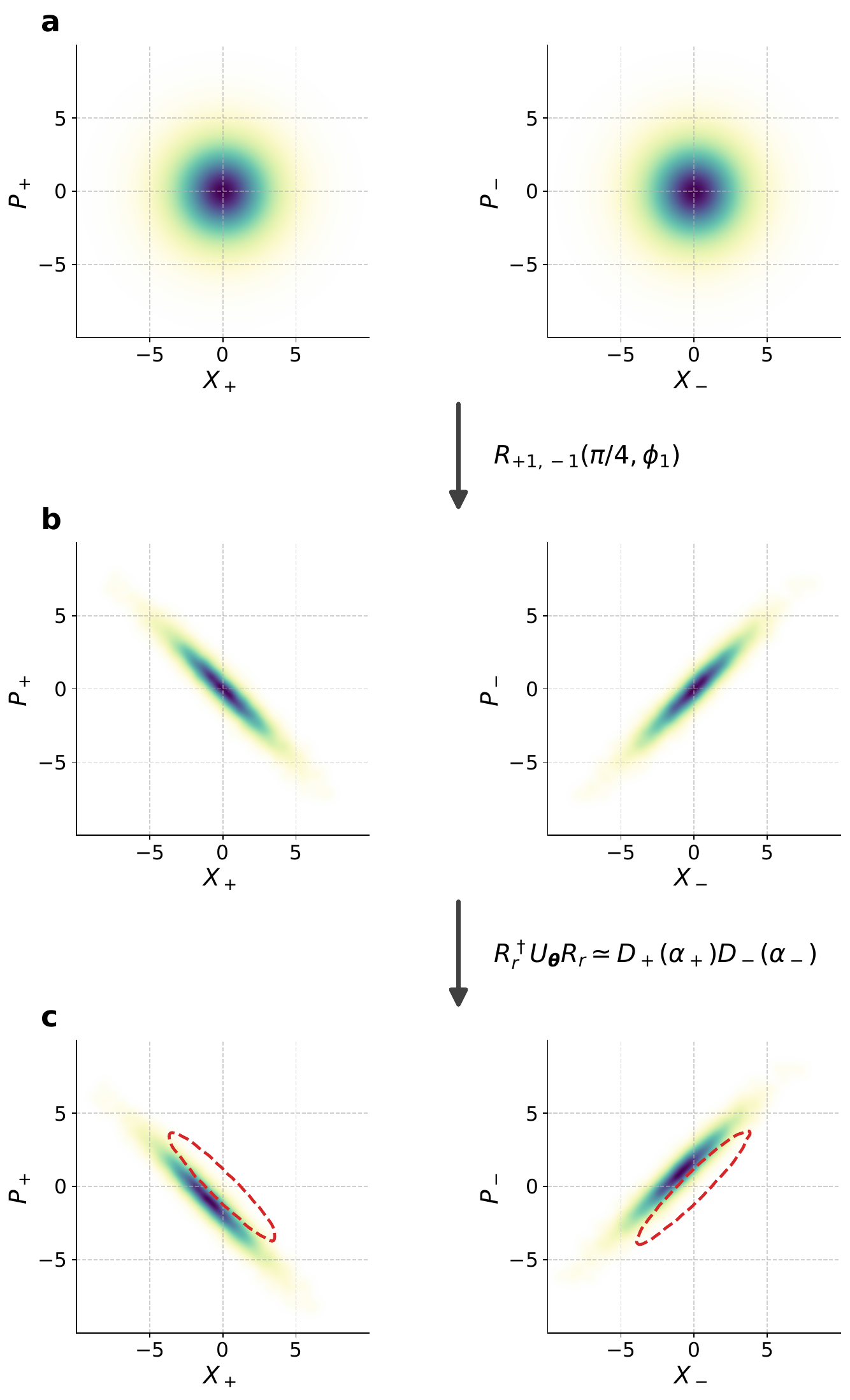}
\caption{Phase-space illustration of the phase-to-displacement mapping.
The left and right columns correspond to the $m_F=+1$ and $m_F=-1$ modes, respectively.
The dimensionless quadratures are defined as $\hat X_\pm=\hat a_{\pm1}+\hat a_{\pm1}^\dagger$ and $\hat P_\pm=-i(\hat a_{\pm1}-\hat a_{\pm1}^\dagger)$, with unit vacuum variance.
The phases are chosen as $\phi_1=0$ and $\phi_r=\pi/4$.
(a) Single-mode marginal distributions of the two-mode squeezed state generated by spin mixing. Each marginal is circular, while the squeezing resides in intermode correlations.
(b) The balanced rotation $R_{+1,-1}(\pi/4,\phi_1)$ converts the intermode correlations into single-mode squeezing.
(c) Within the displacement approximation, the sequence $R_r^\dagger U_{\boldsymbol{\theta}}R_r$ shifts the distribution centers by $(2\operatorname{Re}\alpha_\pm,2\operatorname{Im}\alpha_\pm)$, with the displacement amplitudes $\alpha_\pm$ defined in the main text.
The illustrated displacements lie along the squeezed axes. Red dashed ellipses mark the undisplaced reference contours.}
\label{fig:phase_space_displacements}
\end{figure}

The first operation, $R_{+1,-1}(\pi/4,\phi_1)$, converts the two-mode squeezed vacuum generated by spin mixing into two single-mode squeezed vacua in the undepleted-pump description. Before this rotation, each side mode has an isotropic marginal distribution, as shown in Fig.~\ref{fig:phase_space_displacements}(a), while the squeezing resides in correlations between the two modes. The corresponding transformation of the side-mode operators is
\begin{equation*}
R_{+1,-1}^\dagger
\begin{pmatrix}
\hat a_{+1}\\
\hat a_{-1}
\end{pmatrix}
R_{+1,-1}
=
\frac{1}{\sqrt2}
\begin{pmatrix}
1&-e^{i\phi_1}\\
e^{-i\phi_1}&1
\end{pmatrix}
\begin{pmatrix}
\hat a_{+1}\\
\hat a_{-1}
\end{pmatrix},
\end{equation*}
where the arguments $(\pi/4,\phi_1)$ of the rotation operators have been suppressed. This mixing transforms the intermode correlations into squeezing of the individual modes, producing the two squeezed distributions in Fig.~\ref{fig:phase_space_displacements}(b). Both states have zero mean, so their phase-space distributions remain centered at the origin.

The remaining sequence, $R_r^\dagger U_{\boldsymbol\theta}R_r$, converts the encoded phases into displacements of these two squeezed states. The collective rotation $R_r$ mixes the macroscopic pump field into both side modes, and $R_r^\dagger$ recombines the modes after phase encoding. At zero encoded phase, the two rotations cancel and recover the undisplaced squeezed states. Nonzero phases change the interference during recombination, leaving residual field amplitudes that shift the distribution centers, as illustrated in Fig.~\ref{fig:phase_space_displacements}(c). For $2\sinh^2r\ll N_t$ and $|\theta_\pm|\ll1$, we retain the zero-phase fluctuation operators and the pump-enhanced displacement terms linear in the phases, obtaining
\begin{equation}
\begin{pmatrix}
\hat a_{+1}^{\mathrm{out}}\\
\hat a_{-1}^{\mathrm{out}}
\end{pmatrix}
\simeq
\frac{1}{\sqrt2}
\begin{pmatrix}
1&-e^{i\phi_1}\\
e^{-i\phi_1}&1
\end{pmatrix}
\begin{pmatrix}
\hat a_{+1}\\
\hat a_{-1}
\end{pmatrix}
+
\begin{pmatrix}
\alpha_+\\
\alpha_-
\end{pmatrix},
\label{eq:displacement_approx}
\end{equation}
where the operators on the right describe the original side modes after spin mixing, and
\begin{subequations}
\label{eq:displacement_amplitudes}
\begin{align}
\alpha_+
&=-i\sqrt2\,cs\mu_0e^{-i\phi_r}
\left(c^2\theta_+-s^2\theta_-\right),\\
\alpha_-
&=-i\sqrt2\,cs\mu_0e^{i\phi_r}
\left(s^2\theta_+-c^2\theta_-\right).
\end{align}
\end{subequations}
Here $\mu_0^2\simeq N_t-2\sinh^2r$. Thus, within this leading displacement approximation, the output consists of two displaced single-mode squeezed states. The displacements encode different linear combinations of $\theta_+$ and $\theta_-$, which are independent when $cs(c^2-s^2)\ne0$. The following subsections examine how atomic homodyne and joint parity measurements extract the phase information carried by these displacements.

\subsection{Atomic homodyne detection}
Using the phase-space convention of Fig.~\ref{fig:phase_space_displacements},
we define the fixed output quadratures as
$\hat X_\pm=\hat a_{\pm1}^{\mathrm{out}}
+(\hat a_{\pm1}^{\mathrm{out}})^\dagger$
and
$\hat P_\pm=-i[\hat a_{\pm1}^{\mathrm{out}}
-(\hat a_{\pm1}^{\mathrm{out}})^\dagger]$,
with unit vacuum variance.
To resolve the phase-dependent displacements, we choose the homodyne observables
\begin{equation}
\begin{aligned}
\hat X_\pm^{\mathrm{hom}}
&=
i\left[
e^{\pm i\phi_r}\hat a_{\pm1}^{\mathrm{out}}
-e^{\mp i\phi_r}
(\hat a_{\pm1}^{\mathrm{out}})^\dagger
\right]\\
&=
\mp\sin\phi_r\,\hat X_\pm
-\cos\phi_r\,\hat P_\pm.
\end{aligned}
\label{eq:homodyne_quadrature}
\end{equation}
Here $\phi_r$ is the coupling phase of $R_r$, and the homodyne
reference phases are chosen accordingly.
These observables select the displacement directions defined by
$\alpha_\pm$, with one quadrature measured in each mode.
Within the displacement approximation, the encoded phases shift
the quadrature means without changing the fluctuations of the
underlying single-mode squeezed states.
The quadrature variances are therefore
\begin{equation}
\operatorname{Var}(\hat X_\pm^{\mathrm{hom}})
\simeq
\cosh(2r)-\sin(2\phi_r+\phi_1)\sinh(2r).
\label{eq:homodyne_quadrature_variance}
\end{equation}
Thus, the noise in the displacement signals depends on the relative
orientation of the displacement directions and the squeezed axes.

When the phase-matching condition
$2\phi_r+\phi_1=\pi/2$ modulo $2\pi$ is satisfied,
both quadrature variances are reduced to
$\operatorname{Var}(\hat X_\pm^{\mathrm{hom}})\simeq e^{-2r}$,
while their mean values are
\begin{subequations}
\label{eq:homodyne_mean_signals}
\begin{align}
\langle\hat X_+^{\mathrm{hom}}\rangle
&\simeq
2\sqrt{2}\,cs\mu_0
\left(c^2\theta_+-s^2\theta_-\right),\\
\langle\hat X_-^{\mathrm{hom}}\rangle
&\simeq
2\sqrt{2}\,cs\mu_0
\left(s^2\theta_+-c^2\theta_-\right).
\end{align}
\end{subequations}
Here $\mu_0$ is the pump amplitude.
The two encoded phases are therefore mapped onto displacements
along the squeezed directions.
As illustrated in Fig.~\ref{fig:phase_space_displacements}(c),
the distribution centers shift along the narrow axes of the
single-mode squeezed distributions, so that the signals are
resolved against fluctuations below the vacuum level.
Measuring the two quadrature means then determines
$\theta_+$ and $\theta_-$ through the two linear relations above,
provided that $cs(c^2-s^2)\ne0$.
The reduced quadrature noise supplies the sensitivity enhancement,
while the conversion from displacement to phase is fixed by
these linear responses.
The resulting total phase variance, defined as the sum of the
estimation variances of $\theta_+$ and $\theta_-$, is
\begin{equation}
\Delta\Theta_{\mathrm{hom}}^2
\simeq
\frac{(c^4+s^4)e^{-2r}}
{4\mu_0^2c^2s^2(c^2-s^2)^2}.
\label{eq:homodyne_total_variance}
\end{equation}
This expression coincides with the leading large-$r$ form of the
QCRB derived in Sec.~\ref{sec:qfim}.

The observables $\hat X_\pm^{\mathrm{hom}}$ can, in principle,
be measured by atomic homodyne detection.
A coherent coupling pulse mixes one of the $m_F=\pm1$ modes
with a macroscopically occupied atomic local oscillator,
converting the selected quadrature into a population
difference~\cite{Oberthaler2011}.
In the present system, the pump can provide this reference,
with the coupling phases chosen to measure these two observables.
Experiments on twin-matter-wave interferometry~\cite{Klempt2011},
entanglement between atomic modes~\cite{Klempt2018},
and spin-nematic sensing~\cite{Mao2023}
provide related advances in the preparation and detection
of squeezed atomic resources.
Simultaneous collective-spin readout~\cite{Oberthaler2019}
and joint estimation of a two-phase spin
rotation~\cite{Cao2025JointEstimation}
further motivate access to both output signals within one
experimental realization.
For the protocol considered here, this requires mode-selective
coupling and phase control for both channels.

\subsection{Joint parity detection}
Although atomic homodyne detection provides direct access to the
displacement signals, simultaneous readout of both channels can
require a carefully controlled sequence of mode-selective microwave
pulses~\cite{Cao2025JointEstimation}.
Recent experiments have demonstrated single-atom-resolved
detection of ultracold atoms~\cite{Pur2023,Hetzel2023,Quensen2026},
allowing number parity to be extracted directly from the measured
mode populations.
These experimental advances motivate us to investigate joint parity
detection as an alternative readout.
We measure the parity of the physical $m_F=\pm1$ output modes after
$\mathcal U_2$.
For the output state in Eq.~\eqref{eq:readout_output_state}, define
\begin{equation}
\Pi_\pm=(-1)^{\hat N_{\pm1}},
\label{eq:parity_operators}
\end{equation}
where $\hat N_{\pm1}=\hat a_{\pm1}^\dagger\hat a_{\pm1}$, and
expectation values are evaluated in
$|\psi_{\boldsymbol\theta}^{\mathrm{out}}\rangle$.
Within the leading displacement approximation in
Eq.~\eqref{eq:displacement_approx}, the output consists of two
single-mode squeezed states displaced by $\alpha_\pm$.
Since parity is proportional to the Wigner function at the
phase-space origin, these displacements change the output parity
signals.
Thus, $\Pi_+$ and $\Pi_-$ probe the two phase-dependent
displacement channels.

Under the phase-matching condition $2\phi_r+\phi_1=\pi/2$ modulo
$2\pi$, the phase-induced displacements lie along the squeezed
quadratures of the two output modes.
Within the leading displacement approximation, the mean parity
in mode $k\in\{+,-\}$ is~\cite{Gerry2000,Seshadreesan2013}
\begin{equation}
\langle\Pi_k\rangle
=\exp\!\left(-2e^{2r}|\alpha_k|^2\right).
\label{eq:parity_expectation_optimal}
\end{equation}
At zero displacement, each mode is a squeezed vacuum containing
only even atom numbers, so its mean parity is unity.
A displacement introduces odd-number components and reduces
the mean parity.
The factor $e^{2r}$ means that a given change in parity requires
a displacement smaller by a factor $e^{-r}$ than in the
unsqueezed case.
The two mean parity signals therefore provide information about
the magnitudes of the two phase-dependent displacements.
However, opposite displacements produce identical parity signals,
so these measurements alone do not determine the displacement
directions along the squeezed axes.
Through the phase-to-displacement relations derived above,
this ambiguity corresponds to the unknown signs of
$c^2\theta_+-s^2\theta_-$ and $s^2\theta_+-c^2\theta_-$.
For local estimation, known phase offsets can establish a calibrated
reference point with nonzero displacements in both modes.
At this reference point, the parity responses have known, nonzero
slopes, so changes in the mean parities determine the directions
of small displacement variations.
The local inversion within a region that does not cross either
zero-displacement line is derived in
Appendix~\ref{app:single_mode_parity}.

Figure~\ref{fig:parity_measurement} shows the two parity signals
as functions of the scaled phases $\theta_\pm\sqrt{N_t}$.
In panel (a), each response map has a bright ridge along which
the corresponding displacement vanishes and the mean parity
equals unity.
The different orientations of these ridges reflect the phase
combinations $c^2\theta_+-s^2\theta_-$ and
$s^2\theta_+-c^2\theta_-$ encoded in $\alpha_+$ and $\alpha_-$,
respectively.
Panel (b) shows the two signals along the cut
$\theta_-\sqrt{N_t}=0.2$.
As $\theta_+$ varies, $\langle\Pi_+\rangle$ passes through
a maximum, whereas $\langle\Pi_-\rangle$ increases throughout
the displayed range.
The local estimation precision is determined by these phase
responses together with the parity fluctuations.
\begin{figure}[t]
\centering
\includegraphics[width=0.95\columnwidth]{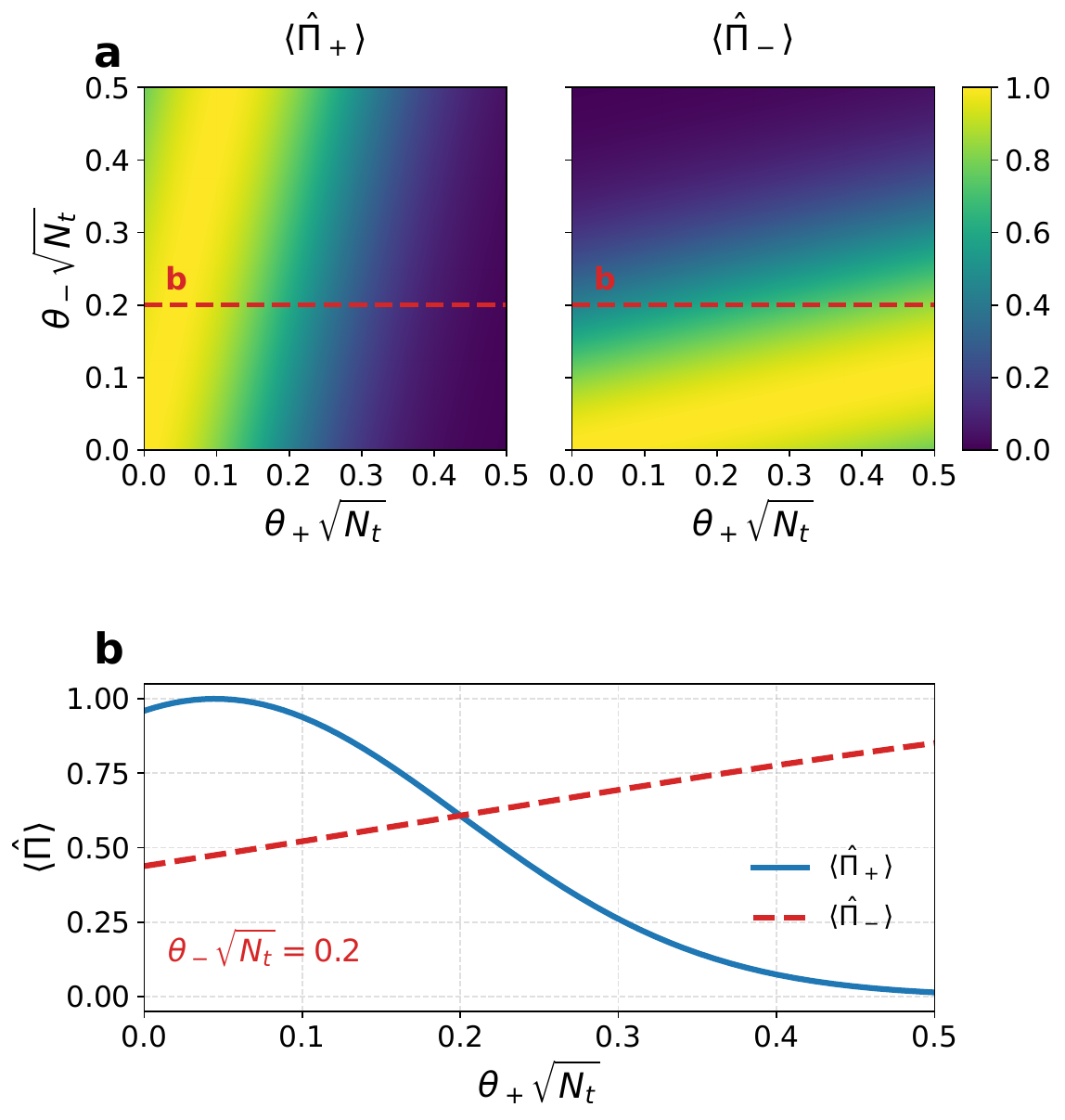}
\caption{Parity signals of the two output modes within the leading
displacement approximation.
(a) Mean parities $\langle\Pi_+\rangle$ (left) and
$\langle\Pi_-\rangle$ (right) as functions of the scaled phases
$\theta_+\sqrt{N_t}$ and $\theta_-\sqrt{N_t}$.
The red horizontal dashed lines indicate the cut
$\theta_-\sqrt{N_t}=0.2$.
(b) Mean parity signals $\langle\Pi_+\rangle$ (blue solid) and
$\langle\Pi_-\rangle$ (red dashed) along this cut as functions
of $\theta_+\sqrt{N_t}$.
Results are calculated from Eq.~\eqref{eq:parity_expectation_optimal}
under phase matching, with $N_t=500$, $r=2.0$,
$\theta_r=0.281\pi$, and $\mu_0=\sqrt{N_t-2\sinh^2r}$.}
\label{fig:parity_measurement}
\end{figure}

\begin{figure}[t]
\centering
\includegraphics[width=0.95\columnwidth]{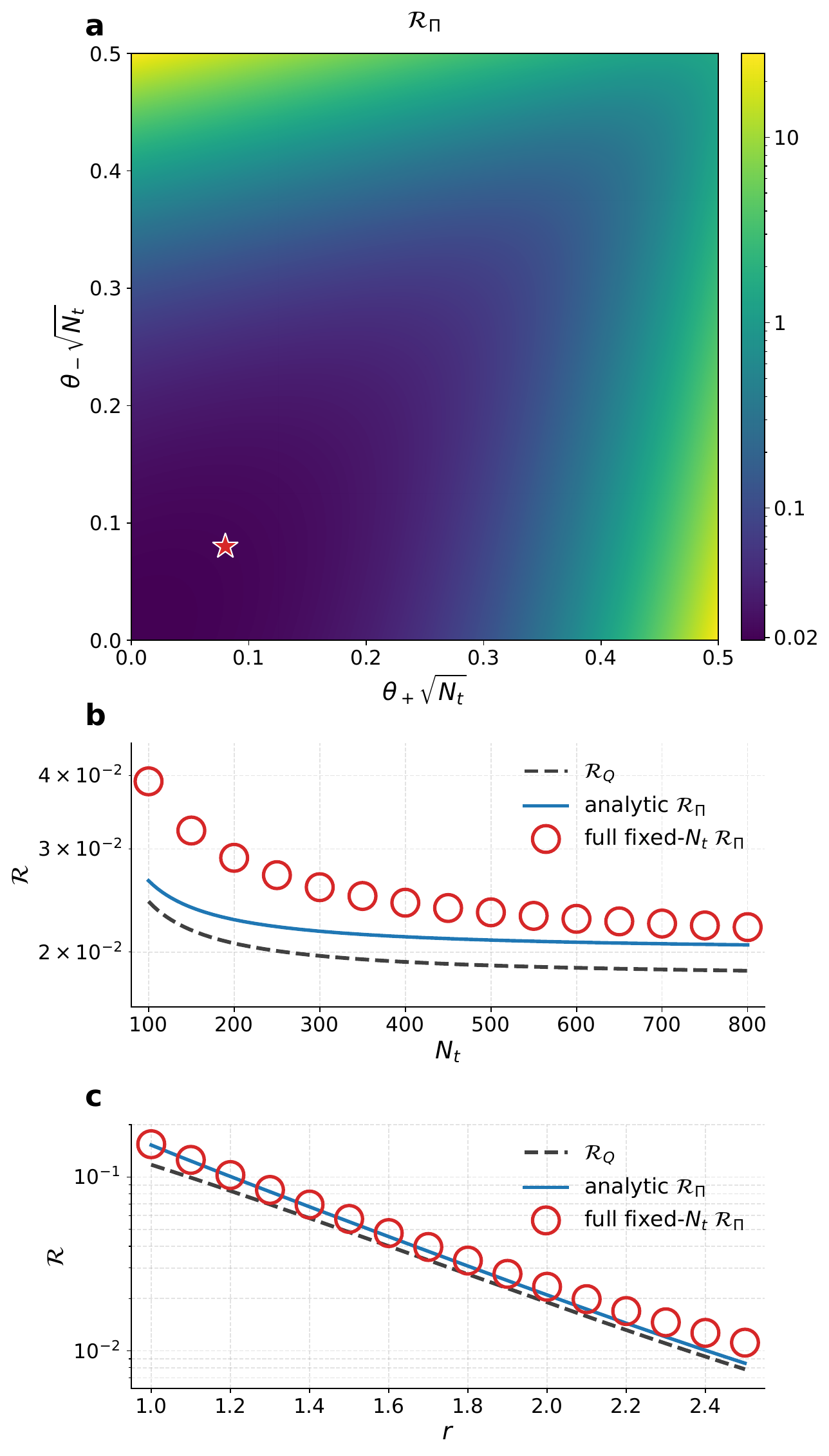}
\caption{SQL-normalized precision of joint parity detection.
All panels use phase matching and the analytical optimal mixing
angle $\theta_r=\theta_r^{\mathrm{opt}}(r)$.
(a) Analytical $\mathcal R_\Pi$ for $N_t=500$ and $r=2.0$, shown on
a logarithmic color scale.
The star marks
$\theta_+\sqrt{N_t}=\theta_-\sqrt{N_t}=0.08$.
(b) Ratios versus $N_t$ at $r=2.0$, with the starred scaled offsets
held fixed.
(c) Ratios versus $r$ for $N_t=500$, with
$\theta_\pm\sqrt{N_t}=0.08\,e^{-(r-2)}$ to keep the analytical
odd-outcome probabilities near $4\%$.
In (b) and (c), gray dashed curves show analytical $\mathcal R_Q$,
blue solid curves show analytical $\mathcal R_\Pi$, and red open
circles show ED $\mathcal R_\Pi$.}
\label{fig:parity_performance}
\end{figure}

We estimate small phase variations about the calibrated operating
point from the mean parity signals obtained by averaging repeated
measurement outcomes.
Each channel returns a parity outcome of $+1$ or $-1$, giving
\begin{equation}
\operatorname{Var}(\Pi_k)=1-\langle\Pi_k\rangle^2.
\end{equation}
Let $\Sigma_n$ denote the covariance matrix of the two parity outcomes,
with elements $(\Sigma_n)_{kl}=\langle\Pi_k\Pi_l\rangle
-\langle\Pi_k\rangle\langle\Pi_l\rangle$ for
$k,l\in\{+,-\}$.
Define the response matrix
$J_{ki}=\partial\langle\Pi_k\rangle/\partial\theta_i$,
with $i\in\{+,-\}$.
All moments and derivatives are evaluated at the operating point,
with the known phase offsets held fixed.
Where $J$ is invertible, the local error-propagation covariance,
reported per interrogation, is
\begin{equation}
\Sigma_\theta^\Pi
=J^{-1}\Sigma_n(J^{-1})^{\mathsf T}.
\label{eq:parity_cov}
\end{equation}
The inverse response converts fluctuations of the measured means
into phase-estimation errors, and the diagonal elements of
$\Sigma_\theta^\Pi$ give the corresponding estimation variances.
Within the leading displacement approximation, the two output side
modes form a product state, so their parity outcomes are independent
and $\Sigma_n=\mathrm{diag}\!\left[\operatorname{Var}(\Pi_+),
\operatorname{Var}(\Pi_-)\right]$.
For the ED results in Fig.~\ref{fig:parity_performance}, we retain
the full covariance matrix to include correlations between the
two parity outcomes.

We quantify the precision at a calibrated nonzero operating point
by the total phase variance,
$\Delta\Theta_\Pi^2=\operatorname{Tr}\Sigma_\theta^\Pi$.
Within the leading displacement approximation, reducing both
displacement magnitudes at fixed probe parameters lowers this
variance.
The zero-displacement limit therefore provides an ideal benchmark
for the finite-offset readout.
This limit is approached through calibrated nonzero reference points
within the local regions specified in
Appendix~\ref{app:single_mode_parity}; at zero displacement itself,
the response matrix $J$ is singular.
The parity variances and squared response slopes vanish at the
same order, yielding, for $cs(c^2-s^2)\ne0$,
\begin{equation}
\lim_{\boldsymbol\theta\to\boldsymbol0}\Delta\Theta_\Pi^2
=
\frac{(c^4+s^4)e^{-2r}}
{4\mu_0^2c^2s^2(c^2-s^2)^2}.
\label{eq:parity_local_total_variance}
\end{equation}
The full limiting covariance in Appendix~\ref{app:single_mode_parity}
coincides with that obtained from the matched homodyne signals and
noise within the same approximation.
At the optimized mixing angle, its trace also agrees with the
leading large-$r$ form of the analytical QCRB in the weak-depletion
regime.

The nonzero reference offsets introduced for local inversion also
serve a statistical purpose.
The odd-outcome probability
\begin{equation}
p_{k,\mathrm{o}}=\frac{1-\langle\Pi_k\rangle}{2},
\qquad k\in\{+,-\},
\label{eq:parity_odd_probability}
\end{equation}
vanishes as the corresponding displacement approaches zero.
Smaller offsets improve the ideal precision per interrogation but
require more repetitions to sample the rare odd outcomes.
For a finite measurement sample, the operating point must therefore
balance proximity to the limiting variance against sufficient
odd-event statistics in both channels.
We use
$\mathcal R_\Pi=\Delta\Theta_\Pi^2/\Delta\Theta_{\mathrm{SQL}}^2$,
with the SQL evaluated for the same total atom number and
equal-weight phase cost.
Values $\mathcal R_\Pi<1$ indicate sub-SQL precision, whereas
comparison with $\mathcal R_Q$ assesses proximity to the probe bound.
Figure~\ref{fig:parity_performance}(a) maps $\mathcal R_\Pi$ over the
scaled phase plane.
The starred point,
$\theta_+\sqrt{N_t}=\theta_-\sqrt{N_t}=0.08$ at $N_t=500$ and $r=2.0$,
lies within the sub-SQL region and gives an analytical odd-outcome
probability of approximately $4\%$ in each channel.

Figures~\ref{fig:parity_performance}(b) and
\ref{fig:parity_performance}(c) compare the analytical and ED parity
ratios with the analytical optimized QCRB ratio $\mathcal R_Q$.
The ED calculations use the full three-mode output state and retain
the complete two-channel parity covariance.
In panel (b), the squeezing and scaled phase offsets are held fixed;
the ED results approach the displacement-model prediction as $N_t$
increases, consistent with decreasing relative pump depletion.
In panel (c), $N_t=500$ and the offsets are chosen as
$\theta_+\sqrt{N_t}=\theta_-\sqrt{N_t}=0.08\,e^{-(r-2)}$.
This scaling follows the narrowing parity response and keeps the
analytical odd-outcome probabilities approximately within
$3.5\%$--$4\%$.
The squeezing scan thus compares different local operating points
with similar odd-event rates, and both parity ratios decrease over
the displayed range.
The separation between the analytical parity and QCRB curves reflects
both the finite-displacement penalty and the finite-$r$ difference
between the local limiting variance and the bound.
Deviations between ED and analytical parity results reflect
corrections beyond the displacement approximation.
These results support local sub-SQL joint phase readout at finite
calibrated operating points.
The benchmarks assume ideal parity resolution; detection loss,
parity misclassification, and global phase unwrapping are not
addressed.

\section{Conclusion}
\label{sec:conclusion}
We have developed a protocol for joint estimation of two relative
phases in a spin-1 Bose--Einstein condensate using correlations
generated by spin mixing.
The pair-correlated probe initially has a rank-one QFIM and carries
only common-phase information.
A parameter-independent passive transformation before encoding
reorganizes these correlations and couples the macroscopic
$m_F=0$ reference to both sensing modes, yielding a full-rank QFIM.
In the undepleted-pump regime, optimization over passive three-mode
transformations reduces to phase matching and a single mixing angle.
An optimal transformation can be implemented as a balanced rotation
between the $m_F=\pm1$ modes followed by a collective spin-1 rotation.
Over the parameter range considered, the optimized equal-weight QCRB
lies below the SQL at the same total atom number $N_t$, with an
enhancement that increases with squeezing.
ED calculations support the analytical predictions in this regime
and quantify deviations from the approximation.

For the optimized, phase-matched probe, a parameter-independent
readout maps two linearly independent phase combinations onto
displacements along squeezed quadratures within the leading
displacement approximation.
These signals can be accessed through atomic homodyne detection
or number-resolved parity measurements.
Calibrated nonzero reference displacements enable local inversion
of the parity signals and provide nonzero odd-outcome probabilities,
at the cost of a precision penalty relative to the zero-displacement
limit.
Both analytical and ED calculations give sub-SQL parity sensitivity
at selected finite operating points.
In the zero-displacement limit, the analytical parity covariance
coincides with the leading homodyne covariance.
Its trace agrees with the leading large-$r$ form of the analytical
QCRB in the weak-depletion regime.

These results provide a route to joint sensing of common and
differential phase shifts associated with quadratic and linear
Zeeman contributions, respectively.
Experimental advances in number-resolved detection and detector
characterization~\cite{Pur2023,Hetzel2023} motivate extending the
parity analysis to calibrated detection models.
Including atom loss and atom-number misclassification would test
how much of the predicted advantage survives under realistic
detection.
Finite-sample estimation would further clarify how the reference
offsets and number of repetitions should be chosen for reliable
local estimation.

\section*{Acknowledgements}
This work is supported by the
National Key R\&D Program of China (Grant
No. 2025YFA1411600), by the National Natural Science Foundation of China (NSFC)
(Grants No.~12304543, No.~92476205, No.~12361131576 and No.~92265205),
and by the Quantum Science and Technology--National Science and Technology Major Project
(Grant No.~2021ZD0302104).

\appendix
\section{Passive Transformations and QFIM Optimization}
\label{app:technical}

\subsection{Matrix Representations of the Passive Transformations}
\label{app:su3_elements}

All single-particle matrices below use the ordered mode basis
$(m_F=0,+1,-1)$.
The diagonal phase matrix is
$D(\delta_1, \delta_2) = \text{diag}(e^{-i(\delta_1+\delta_2)}, e^{i\delta_1}, e^{i\delta_2})$.
The embedded $SU(2)$ rotations are
\begin{align}
R_{+1,-1}(\theta_1, \phi_1) &=
\begin{pmatrix}
1 & 0 & 0 \\ 0 & \cos\theta_1 & -e^{i\phi_1}\sin\theta_1 \\ 0 & e^{-i\phi_1}\sin\theta_1 & \cos\theta_1
\end{pmatrix}, \\
R_{0,+1}(\theta_2, \phi_2) &=
\begin{pmatrix}
\cos\theta_2 & -e^{i\phi_2}\sin\theta_2 & 0 \\ e^{-i\phi_2}\sin\theta_2 & \cos\theta_2 & 0 \\ 0 & 0 & 1
\end{pmatrix}, \\
R_{0,-1}(\theta_3, \phi_3) &=
\begin{pmatrix}
\cos\theta_3 & 0 & -e^{i\phi_3}\sin\theta_3 \\ 0 & 1 & 0 \\ e^{-i\phi_3}\sin\theta_3 & 0 & \cos\theta_3
\end{pmatrix}.
\end{align}
For the factorization in Eq.~\eqref{eq:su3_factorization},
the single-particle matrix $U$ associated with $\mathcal U_1$
has elements
\begin{align}
U_{11}&=e^{-i(\delta_1+\delta_2)} \cos\theta_3 \cos\theta_2 \label{eq:U_11}\\
U_{12}&=-e^{-i(\delta_1+\delta_2)}
\left[
\begin{aligned}
&e^{i\phi_2}\cos\theta_3\sin\theta_2\cos\theta_1\\
&{}+e^{i(\phi_3-\phi_1)}\sin\theta_3\sin\theta_1
\end{aligned}
\right]\label{eq:U_12}\\
U_{13}&=e^{-i(\delta_1+\delta_2)}
\left[
\begin{aligned}
&e^{i(\phi_2+\phi_1)}\cos\theta_3\sin\theta_2\sin\theta_1\\
&{}-e^{i\phi_3}\sin\theta_3\cos\theta_1
\end{aligned}
\right]\label{eq:U_13}\\
U_{21}&=e^{i(\delta_1-\phi_2)} \sin\theta_2 \label{eq:U_21}\\
U_{22}&=e^{i\delta_1} \cos\theta_2 \cos\theta_1 \label{eq:U_22}\\
U_{23}&=-e^{i(\delta_1+\phi_1)} \cos\theta_2 \sin\theta_1 \label{eq:U_23}\\
U_{31}&=e^{i(\delta_2-\phi_3)} \sin\theta_3 \cos\theta_2 \label{eq:U_31}\\
U_{32}&=e^{i\delta_2} [ e^{-i\phi_1} \cos\theta_3 \sin\theta_1 - e^{i(\phi_2-\phi_3)} \sin\theta_3 \sin\theta_2 \cos\theta_1 ] \label{eq:U_32}\\
U_{33}&=e^{i\delta_2} [ \cos\theta_3 \cos\theta_1 + e^{i(\phi_1+\phi_2-\phi_3)} \sin\theta_3 \sin\theta_2 \sin\theta_1 ] \label{eq:U_33}
\end{align}

The matrix of the collective spin-$1$ rotation $R_r(\theta_r,\phi_r)$, with $c=\cos(\theta_r/2)$ and $s=\sin(\theta_r/2)$, is given in Eq.~\eqref{eq:Rr_definition}.

Here $U'$ denotes the single-particle matrix of the
structured pre-encoding transformation $\mathcal U_1$
in Eq.~\eqref{eq:structured_U}, whereas $U$ above denotes
the matrix in its general parameterization.
The elements in the second and third columns of $U'$ are
\begin{align}
U'_{12}&=-cs e^{i\phi_r} + cs e^{-i(\phi_r+\phi_1)} \label{eq:U'_12}\\
U'_{13}&= cs e^{i(\phi_r+\phi_1)} + cs e^{-i\phi_r} \label{eq:U'_13}\\
U'_{22}&=\frac{\sqrt{2}}{2}c^2 + \frac{\sqrt{2}}{2} s^2 e^{-i (2\phi_r+\phi_1)} \label{eq:U'_22}\\
U'_{23}&=-\frac{\sqrt{2}}{2}c^2e^{i\phi_1}+ \frac{\sqrt{2}}{2} s^2 e^{-i 2\phi_r} \label{eq:U'_23}\\
U'_{32}&=\frac{\sqrt{2}}{2}c^2e^{-i\phi_1} + \frac{\sqrt{2}}{2} s^2 e^{i 2\phi_r} \label{eq:U'_32}\\
U'_{33}&=\frac{\sqrt{2}}{2}c^2-\frac{\sqrt{2}}{2} s^2 e^{i (2\phi_r+\phi_1)} \label{eq:U'_33}
\end{align}

\subsection{Generator Covariances and the General QFIM}
\label{app:noise_stats}

This subsection evaluates the generator covariances used in Sec.~III.B.
Let $\hat a'_\pm\equiv\hat a_{\pm1}(t)$ denote the side-mode
operators after spin mixing, as given by Eq.~\eqref{eq:a_pm_Bog}.
For each sensing channel $m\in\{+,-\}$, write the effective mode
and its phase generator as
\begin{equation}
\hat{\widetilde a}_m=\Gamma_m\hat a_0+\hat\gamma_m,
\qquad
\hat G_m=\hat{\widetilde a}_m^\dagger\hat{\widetilde a}_m,
\label{eq:app_effective_mode_decomposition}
\end{equation}
where $(\Gamma_+,\Gamma_-)=(U_{21},U_{31})$ and
\begin{equation}
\hat\gamma_+=U_{22}\hat a'_+ +U_{23}\hat a'_-,
\qquad
\hat\gamma_-=U_{32}\hat a'_+ +U_{33}\hat a'_-.
\label{eq:app_noise_operators}
\end{equation}
Within the analytical approximation, we treat the pump as a coherent
reference with $\operatorname{Var}(\hat N_0)=\mu_0^2$ and assume that
its fluctuations are independent of the side-mode noise.
Retaining terms of order $\mu_0^2$ gives
\begin{align}
\operatorname{Cov}(\hat G_m,\hat G_n)
\simeq{}&\mu_0^2\Big[
|\Gamma_m|^2|\Gamma_n|^2
\nonumber\\
&+\Gamma_m^*\Gamma_n^*\langle\hat\gamma_m\hat\gamma_n\rangle
+\Gamma_m^*\Gamma_n\langle\hat\gamma_m\hat\gamma_n^\dagger\rangle
\nonumber\\
&+\Gamma_m\Gamma_n^*\langle\hat\gamma_m^\dagger\hat\gamma_n\rangle
+\Gamma_m\Gamma_n\langle\hat\gamma_m^\dagger\hat\gamma_n^\dagger\rangle
\Big].
\label{eq:cov_approx_appendix}
\end{align}
The QFIM elements used in the main text are $F_{mn}=4\operatorname{Cov}(\hat G_m,\hat G_n)$.

The Bogoliubov side-mode moments are
\begin{align}
\langle (\hat a'_\pm)^\dagger\hat a'_\pm\rangle&=\sinh^2r,
&
\langle\hat a'_\pm(\hat a'_\pm)^\dagger\rangle&=\cosh^2r,
\nonumber\\
\langle\hat a'_+\hat a'_-\rangle&=\frac{i}{2}\sinh2r,
&
\langle(\hat a'_+)^\dagger(\hat a'_-)^\dagger\rangle&=-\frac{i}{2}\sinh2r.
\label{eq:app_bogoliubov_moments}
\end{align}
The normal and anomalous noise correlations follow from
Eqs.~\eqref{eq:app_noise_operators}
and~\eqref{eq:app_bogoliubov_moments}.
With row indices $r_+=2$ and $r_-=3$, they are
\begin{align}
\langle\hat\gamma_m^\dagger\hat\gamma_n\rangle
&=(U_{r_m2}^*U_{r_n2}+U_{r_m3}^*U_{r_n3})\sinh^2r,
\nonumber\\
\langle\hat\gamma_m\hat\gamma_n^\dagger\rangle
&=(U_{r_m2}U_{r_n2}^*+U_{r_m3}U_{r_n3}^*)\cosh^2r,
\nonumber\\
\langle\hat\gamma_m\hat\gamma_n\rangle
&=\frac{i}{2}(U_{r_m2}U_{r_n3}+U_{r_m3}U_{r_n2})\sinh2r,
\nonumber\\
\langle\hat\gamma_m^\dagger\hat\gamma_n^\dagger\rangle
&=-\frac{i}{2}(U_{r_m2}^*U_{r_n3}^*+U_{r_m3}^*U_{r_n2}^*)\sinh2r.
\label{eq:app_noise_correlations}
\end{align}
Substituting these correlations into
Eq.~\eqref{eq:cov_approx_appendix} for the factorization in
Eq.~\eqref{eq:su3_factorization} gives the leading-order
QFIM elements $A$, $B$, and $C$,
\begin{widetext}
\begin{align}
A \simeq{}&4\mu_0^2\sin^4\theta_2
+\mu_0^2\sin^22\theta_2
\left(\cosh2r+\sin2\theta_1\sin\Phi_3\sinh2r\right),
\label{eq:A_general}\\
B \simeq{}&4\mu_0^2\sin^4\theta_3\cos^4\theta_2
+4\mu_0^2\sin^2\theta_3\cos^2\theta_2
\Big[(\cos^2\theta_3+\sin^2\theta_3\sin^2\theta_2)\cosh2r
\nonumber\\
&+\big(-\cos^2\theta_3\sin2\theta_1\sin\Phi_2
+\sin^2\theta_3\sin^2\theta_2\sin2\theta_1\sin\Phi_3
+\sin2\theta_3\sin\theta_2\cos2\theta_1\sin\Phi_1\big)\sinh2r\Big],
\label{eq:B_general}\\
C \simeq{}&\mu_0^2\sin^2\theta_3\sin^22\theta_2
-\mu_0^2\sin\theta_3\sin2\theta_2
\Big[\sin\theta_3\sin2\theta_2\cosh2r
\nonumber\\
&+\big(2\cos\theta_3\cos\theta_2\cos2\theta_1\sin\Phi_1
+\sin\theta_3\sin2\theta_2\sin2\theta_1\sin\Phi_3\big)\sinh2r\Big].
\label{eq:C_general}
\end{align}
\end{widetext}
The phase combinations are
\begin{equation}
\Phi_1=\phi_2+\phi_3,
\qquad
\Phi_2=-\phi_1+2\phi_3,
\qquad
\Phi_3=\phi_1+2\phi_2,
\label{eq:app_general_phase_combinations}
\end{equation}
with $2\Phi_1=\Phi_2+\Phi_3$. The corresponding equal-weight QCRB is
\begin{equation}
\Delta\Theta_{\mathrm{QCRB}}^2
=\operatorname{Tr}(F_Q^{-1})
=\frac{A+B}{AB-C^2}.
\label{eq:app_general_qcrb}
\end{equation}
Under the optimal phase conditions
\begin{equation}
\Phi_2=\left(2k_2-\frac12\right)\pi,
\qquad
\Phi_3=\left(2k_3+\frac12\right)\pi,
\qquad
k_2,k_3\in\mathbb Z,
\label{eq:opt_phase_condition}
\end{equation}
These conditions imply $\sin\Phi_1=0$, so
Eqs.~\eqref{eq:A_general}--\eqref{eq:C_general} reduce to
\begin{align}
A \simeq{}&4\mu_0^2\sin^4\theta_2
+\mu_0^2\sin^22\theta_2
\left(\cosh2r+\sin2\theta_1\sinh2r\right),
\label{eq:A_opt}\\
B \simeq{}&4\mu_0^2\sin^4\theta_3\cos^4\theta_2
+4\mu_0^2\sin^2\theta_3\cos^2\theta_2
\nonumber\\
&\times(\cos^2\theta_3+\sin^2\theta_3\sin^2\theta_2)
\left(\cosh2r+\sin2\theta_1\sinh2r\right),
\label{eq:B_opt}\\
C \simeq{}&\mu_0^2\sin^2\theta_3\sin^22\theta_2
\left[1-\left(\cosh2r+\sin2\theta_1\sinh2r\right)\right].
\label{eq:C_opt}
\end{align}
Equations~\eqref{eq:A_general}--\eqref{eq:app_general_qcrb}
specify the QFIM and the corresponding equal-weight QCRB
for the general transformation used in Sec.~III.B.

\subsection{Phase Matching and Optimal Mixing}
\label{app:squeezing_analysis}

For the structured transformation in Eq.~\eqref{eq:structured_U},
the pump-coupling amplitudes are
\begin{equation}
\Gamma_+=\sqrt2\,cs e^{-i\phi_r},
\qquad
\Gamma_-=-\sqrt2\,cs e^{i\phi_r}.
\label{eq:app_structured_pump_couplings}
\end{equation}
Substituting the matrix elements in Appendix~\ref{app:su3_elements}
into Eq.~\eqref{eq:cov_approx_appendix} gives
\begin{equation}
\begin{aligned}
A=B\simeq{}&16\mu_0^2c^4s^4
+8\mu_0^2c^2s^2(c^4+s^4)\chi(\delta\Phi),\\
C\simeq{}&16\mu_0^2c^4s^4
-16\mu_0^2c^4s^4\chi(\delta\Phi),
\end{aligned}
\label{eq:app_ABC_simplified}
\end{equation}
where
\begin{equation}
\delta\Phi=2\phi_r+\phi_1-\frac{\pi}{2},
\qquad
\chi(\delta\Phi)=\cosh2r+\cos(\delta\Phi)\sinh2r.
\label{eq:app_chi_definition}
\end{equation}
Equivalently,
$\chi=\cosh2r+\sin(2\phi_r+\phi_1)\sinh2r$.
The phase-matching condition $2\phi_r+\phi_1=\pi/2$ therefore sets
$\delta\Phi=0$ and gives $\chi=e^{2r}$.

At phase matching, substituting
Eq.~\eqref{eq:app_ABC_simplified} into
$\operatorname{Tr}(F_Q^{-1})$ gives the angle-dependent cost in
Eq.~\eqref{eq:qcrb_Uprime}.
Its minimum equals that obtained by optimizing the general
functional in Eq.~\eqref{eq:app_general_qcrb} over the full passive
transformation space.
The remaining freedom in a general optimizer changes only
unmeasured phases or the choice of an equivalent squeezed-noise
basis, leaving the QFIM for the two encoded phases unchanged.

On the large-squeezing branch, minimizing
Eq.~\eqref{eq:qcrb_Uprime} with respect to $\theta_r$ gives the
optimal angle in Eq.~\eqref{eq:optimal_angle_large_r}.
Substituting this angle yields the minimum QCRB in
Eq.~\eqref{eq:minimum_qcrb_large_r}.
For comparison with ED in Sec.~III.B, we use the approximate
population relation $\mu_0^2\simeq N_t-2\sinh^2r$.

\section{Local Inversion and Covariance for Joint Parity Detection}
\label{app:single_mode_parity}

We use the leading displacement approximation at phase matching
to describe local inversion about a calibrated reference point
and derive the limiting covariance in Eq.~\eqref{eq:parity_local_cov}.
The real, signed displacements are
\begin{equation}
\begin{aligned}
d_+&=\sqrt2\,cs\mu_0(c^2\theta_+-s^2\theta_-),\\
d_-&=\sqrt2\,cs\mu_0(s^2\theta_+-c^2\theta_-).
\end{aligned}
\label{eq:app_parity_displacements}
\end{equation}
Equations~\eqref{eq:displacement_amplitudes}
and~\eqref{eq:homodyne_mean_signals} give $|\alpha_k|^2=d_k^2$
and $\langle\hat X_k^{\mathrm{hom}}\rangle\simeq2d_k$, respectively.
Thus $d_k$ is the signed displacement along the measured squeezed
quadrature.
Equation~\eqref{eq:parity_expectation_optimal} becomes
$\langle\Pi_k\rangle=\exp(-2e^{2r}d_k^2)$, which depends only on
$d_k^2$ and is insensitive to its sign. 

To select a local sign branch, let $\boldsymbol\beta$ denote
the known reference phases set by controlled phase offsets.
The total encoded phases and displacements are
\begin{equation}
\begin{aligned}
\boldsymbol\theta&=\boldsymbol\beta+\delta\boldsymbol\theta,\\
d_k&=d_{k0}+\delta d_k,
\end{aligned}
\label{eq:app_parity_reference_point}
\end{equation}
where $d_{k0}=d_k(\boldsymbol\beta)$ and
$\delta d_k=d_k(\delta\boldsymbol\theta)$ follow from the linear
relations in Eq.~\eqref{eq:app_parity_displacements}.
Choose $d_{k0}\ne0$ in both modes and restrict the unknown
variations to
\begin{equation}
|\delta d_k|<|d_{k0}|,\qquad k\in\{+,-\}.
\label{eq:app_parity_local_branch}
\end{equation}
Each displacement then retains the sign of its calibrated
reference value.

The mean parity at the reference point is
$\langle\Pi_k\rangle_0=\exp(-2e^{2r}d_{k0}^2)$.
For small variations, the change
$\delta\langle\Pi_k\rangle=\langle\Pi_k\rangle-\langle\Pi_k\rangle_0$
satisfies
\begin{equation}
\delta\langle\Pi_k\rangle
=-4e^{2r}d_{k0}\langle\Pi_k\rangle_0\,\delta d_k
+O(\delta d_k^2).
\label{eq:app_parity_reference_response}
\end{equation}
The response slope is nonzero and has a known sign, so changes
in mean parity determine the direction of the displacement variation.
Inverting the linear relations in
Eq.~\eqref{eq:app_parity_displacements} then recovers the two
phase variations for $cs(c^2-s^2)\ne0$.
The zero-displacement limit is approached through nonzero reference
points, with the local region at each point restricted by
Eq.~\eqref{eq:app_parity_local_branch}.

Within the leading displacement approximation, the two output
channels are independent, so $\Sigma_n$ is diagonal.
Using the response matrix $J$, the inverse of the covariance in
Eq.~\eqref{eq:parity_cov} can therefore be written as
\begin{equation}
\left(\Sigma_\theta^\Pi\right)^{-1}_{ij}
=\left(J^{\mathsf T}\Sigma_n^{-1}J\right)_{ij}
=\sum_{k=\pm}
\frac{\partial_i\langle\Pi_k\rangle\partial_j\langle\Pi_k\rangle}
{1-\langle\Pi_k\rangle^2}.
\label{eq:app_parity_cov_inverse}
\end{equation}
At zero displacement, direct substitution in each term gives $0/0$.
Expanding the parity signal for small nonzero $d_k$ gives
\begin{equation}
\begin{aligned}
\partial_i\langle\Pi_k\rangle
&=-4e^{2r}d_k\,\partial_i d_k+O(d_k^3),\\
1-\langle\Pi_k\rangle^2
&=4e^{2r}d_k^2+O(d_k^4).
\end{aligned}
\label{eq:app_parity_expansion}
\end{equation}
Hence each channel contributes the finite limit
\begin{equation}
\lim_{d_k\to0}
\frac{\partial_i\langle\Pi_k\rangle\partial_j\langle\Pi_k\rangle}
{1-\langle\Pi_k\rangle^2}
=4e^{2r}(\partial_i d_k)(\partial_j d_k),
\label{eq:app_parity_regular_limit}
\end{equation}
where the derivatives of $d_k$ follow from
Eq.~\eqref{eq:app_parity_displacements}.
Summing the contributions from both channels gives
\begin{equation}
\left(\Sigma_\theta^\Pi\right)^{-1}\to
8e^{2r}\mu_0^2c^2s^2
\begin{pmatrix}
c^4+s^4&-2c^2s^2\\
-2c^2s^2&c^4+s^4
\end{pmatrix}.
\label{eq:app_parity_local_cov}
\end{equation}
For $cs(c^2-s^2)\ne0$, inverting this matrix gives
\begin{equation}
\lim_{\boldsymbol\theta\to\boldsymbol0}\Sigma_\theta^\Pi
=\frac{e^{-2r}}{8\mu_0^2c^2s^2(c^2-s^2)^2}
\begin{pmatrix}
c^4+s^4&2c^2s^2\\
2c^2s^2&c^4+s^4
\end{pmatrix}.
\label{eq:parity_local_cov}
\end{equation}
The trace of this covariance gives
Eq.~\eqref{eq:parity_local_total_variance}.
This limit applies within a known sign branch because the parity
signal is even in $d_k$.

\end{document}